# Sector Volatility Prediction Performance Using GARCH Models and Artificial Neural Networks

October 1, 2020

**Curtis Nybo**
University of London
nybo.curtis@gmail.com

## Abstract

Recently neural networks (ANNs) have seen success in volatility prediction, but the literature is divided on where an ANN should be used rather than the common GARCH model. The purpose of this study is to compare the volatility prediction performance of ANN and GARCH models when applied to stocks with low, medium, and high volatility profiles. This approach intends to identify which model should be used for each case. The volatility profiles comprise of five sectors that cover all stocks in the U.S stock market from 2005 to 2020. Three GARCH specifications and three ANN architectures are examined for each sector, where the most adequate model is chosen to move on to forecasting. The results indicate that the ANN model should be used for predicting volatility of assets with low volatility profiles, and GARCH models should be used when predicting volatility of medium and high volatility assets.

## 1 Introduction

A prominent feature of capital markets is the uncertainty around price movements. This is referred to as volatility, which measures the dispersion of an assets returns around its average, which in turn acts as a measure of risk. Volatility is a latent variable that cannot be observed at an instantaneous point in time. Therefore, we resort to using averaging methods as a proxy for the true volatility value such as the common variance or standard deviation measure. Riskier assets often have large differences between the current and average price over a period of time, while less risky asset prices tend to remain closer to their average price. Since an increase in volatility leads to an increase in risk, it similarly leads to an increase in potential profits given by the larger dispersion of returns. This phenomenon makes volatility a popular topic of study in the finance literature.

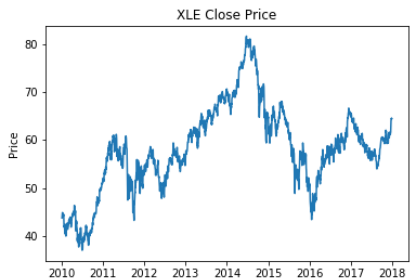
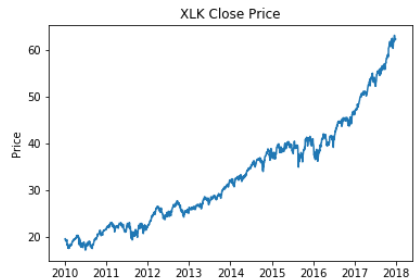
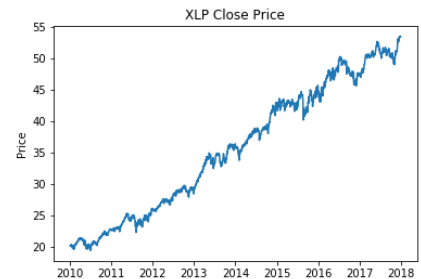

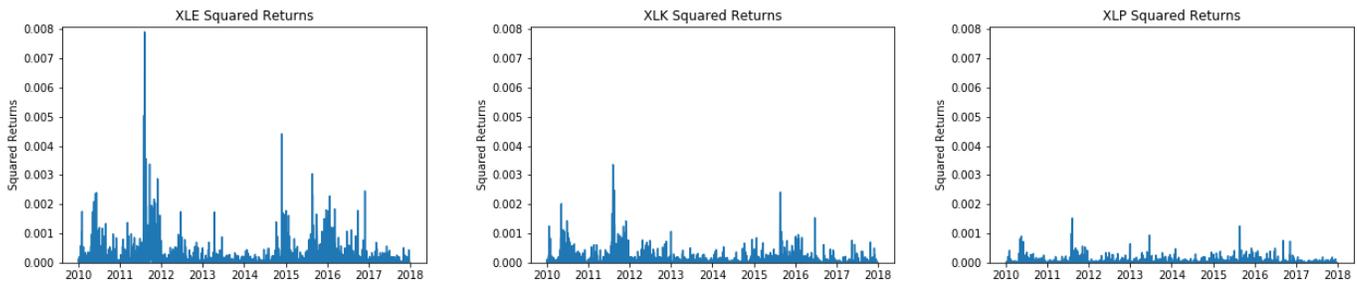

*Figure 1. Three sectors with different volatility profiles, represented by sector ETFs (Yahoo Finance, 2020)*

The importance of volatility as a risk measure is the main reason for incorporating it into financial analysis. Every stock faces unique risks that pertain to its own operations or line of business, and this risk is visible in the up and down movement of stock prices. Figure 1 displays three example sectors that have different levels of volatility. Although these charts merely depict prices and trends, price charts are often a market participant's first indication of an asset's volatility. The XLE ETF represents the very volatile energy sector, as seen in its wild price movements with very little indication of an overall trend. The XLK ETF pertains to the technology sector, which has much less volatility with very few price swings and an overall upward trend is evident. The XLP ETF tracks the consumer staples sector and as such has the lowest volatility of the three examples. The trend is overall quite smooth with very few volatility shocks. This illustrates the requirement for each stock or asset to be analyzed with a model that properly tracks the risks and level of price variation inherent to that particular stock.

By capturing volatility accurately, risk can be quantified and forecasted in all areas of finance. For example, in portfolio management and trading, volatility is an important variable in determining asset allocation criteria. The aim of the risk management is to help a portfolio or investment achieve a specific level of performance, while limiting the probability of negative performance. Volatility even plays a prominent role in not only how risk managers measure risk, but also how they minimize it. Derivative securities are often used to limit the exposure of asset or portfolio performance to market volatility. In order to accurately price derivative securities for risk management, the same measure of volatility is used to first quantify the portfolio risk, then to provide an accurate and fair price for the derivative intended to minimize that risk. Volatility plays an even larger role in finance and economics than the examples given at the individual or institutional level. Market volatility has a large influence on economic factors and country-level policy decisions. Economic policy uncertainty was found by Liu and Zhang (2015) to lead to an increase in stock market volatility, while Krol (2014) found a similar result where U.S. economic policy uncertainty led to an increase in volatility in several different currency exchange rates.

Linear models are the workhorse of many forecasting models as they excel at forecasting time-series data with linear features and are very simple and intuitive to understand in many cases. However, volatility data contains the non-linear features of leptokurtosis, clustering, and leverage effects (Brooks, 2014). Leptokurtosis refers to the increased probability of volatility data containing extreme outliers. Volatility clustering is the tendency of large changes in asset returns to be followed by more large changes. This can be identified where high volatility persists for some time before returning to a lower, normal level of volatility (e.g. stock market crashes). The leverage effect is a well-known phenomenon in finance, where a negative price movement results in higher volatility than a positive move of the same magnitude. This is often attributed to an increase in the debt/equity ratio when a stock price moves lower, thereby increasing the level of debt exposure when the stock price drops thereby further increasing the risk (default risk), while the opposite is true for a rise in the stock price. These features are inherent in stock volatility data and require specific non-linear methods to accurately model and forecast volatility values.

The Autoregressive Conditional Heteroscedastic (ARCH) model was developed to deal with these non-linear issues (Engle, 1982). The model can accommodate autoregressive behaviour such as volatility clustering. Since autoregressive behaviour results in a variance that is dependant, or conditional, on the prior period variance, these variances are then non-constant over a period of time. This condition is known as heteroscedasticity, which causes imprecise confidence intervals and therefore poses a challenge for common statistical forecasting techniques. Given the labels in the ARCH acronym, this model was developed to address these issues. The ARCH model is widely used in practice but suffers from issues such as determining the appropriate number of lags and non-negativity constraints. Bollerslev (1986) built upon the work of Engle and developed the Generalized ARCH (GARCH) model, which addresses many issues of and improves upon the ARCH model. This has led to the GARCH model often being chosen as the appropriate model to describe volatility in academics and industry.

The advancements in data science research in the last two decades combined with the decreasing cost of computing power has led artificial intelligence and machine learning models to see widespread applications in finance. The study of volatility prediction has been one of these focuses. The artificial neural network (ANN) has proven effective in solving various prediction problems, from regression and classification to anomaly detection. The application of ANN models to volatility predictions has proven to be promising since earlier studies in the field from researchers such as Donaldson and Kamstra (1997). More recently, ANNs have proven to be effective in modelling non-linear data due to their reduced dependence on the assumptions placed on the classical methods. This makes ANNs well suited for volatility prediction and has led to increased research in this area.

The forecasting performance comparisons between GARCH and ANN models in the existing literature focus mainly on prediction of stock market indexes covering all stocks from a certain country (e.g. NYSE, DAX) or a specific stock or commodity. Both models have been shown to outperform one another in different forecasting experiments, with little work completed in the sense of understanding which specific situations each model may be better suited. The approach in this study aims to understand these specific situations where each model may perform best by predicting volatility in specific market sectors with varying levels of volatility.

The objective of this study is twofold. First, we develop a suitable GARCH and ANN model to compare the forecasting performance of each model when applied to specific volatility profiles to identify the conditions where one model may outperform the other. Second, we demonstrate the potential usefulness of ANNs for volatility prediction within these volatility profiles. The volatility profiles are made up of five sector portfolios that contain all stocks traded on the U.S stock exchanges, sorted into their appropriate sector grouping. Examining ANN prediction performance of volatility at the sector level expands upon the existing literature as it narrows down specifically where an ANN model may be most useful in predicting volatility. This study contributes to the current body of knowledge around the circumstances required for ANNs to outperform GARCH models.

The results of this study show that the ANN models outperform the GARCH models when predicting the volatility of low volatility sectors (consumer durables and health). The GARCH models outperform the ANN model when predicting volatility in the medium (technology) and high volatility sectors (manufacturing, and other). This indicates that the ANNs are better suited at predicting volatility in assets with low volatility profiles, while GARCH models are better suited for assets with medium and high volatility profiles.

The paper is structured with a literature review in section 2 detailing the current research in the area of volatility forecasting with GARCH and ANN models. Section 3 provides specifics of the methodology and relevant theories used in the GARCH and ANN model specifications and predictions. The sector data used to represent distinct volatility profiles is detailed in section 4. Section 5 contains the results of the model specification and prediction performance. A discussion of these results and their implications follows in section 6. Section 7 presents a conclusion and suggestions for further research.

# 2  Literature Review

The challenge of developing accurate predictive models has long been a top priority for researchers. The earlier works of the ARCH (Engle, 1982) and GARCH (Bollerslev, 1986) models spearheaded the research of volatility prediction by developing the models specifically to handle the non-linear features of stock return data. The research into the feasibility of these models to predict volatility in practice and academics has since greatly expanded.

Liu, et al. (2018) demonstrated the use of a GARCH model to deliver accurate volatility values for derivative valuation. The GARCH framework used was found to provide lower premiums for a put option on a bank's assets (known more commonly as deposit insurance), in comparison to the common Black-Scholes model. Liu found that since the Black-Scholes model assumes constant volatility resulting in over-estimating the bank default risk during higher risk periods, a GARCH model addresses this issue and results in more accurate derivative premium calculations. Kim and Enke (2016) provide another example where a GARCH(1,1) model was shown to decrease risk in a stock portfolio following a target-volatility asset allocation strategy. The GARCH model is used to predict a level of volatility at a given time, and the proportion of risky vs risk-free assets is adjusted based on whether the predicted volatility value is higher or lower than the target value. This approach resulted in a lower maximum drawdown while maintaining a similar Sharpe ratio when compared to the common fixed asset allocation strategy of 60/40 in risky and risk-free assets, respectively.

GARCH-type models have mainly been studied and applied to predict corporate stock, country stock index, or commodity price volatility. Much research has been completed in testing the various GARCH-type models in these markets using linear vs nonlinear and symmetric vs asymmetric GARCH models. Wei, et al. (2010) built upon prior work to use GARCH models to capture the volatility characteristics of the Brent and West Texas Intermediate (WTI) oil markets. Of the eight GARCH-type models tested, no single model was found to be superior to the rest in all situations. However, the non-linear (e.g. GJR-GARCH, FIGARCH) models in general were found to provide better forecasting accuracy compared to the linear models (e.g. GARCH, IGARCH). This is due to the ability to capture the asymmetry and memory effects of oil prices. A similar approach was applied to the gold market by Bentes (2015), where GARCH($p,q$), IGARCH($p,q$), FIGARCH($p,d,q$) models are applied to gold spot prices. Once again, the non-linear FIGARCH($1,d,1$) model was determined to be the optimal model for gold volatility forecasts, indicating that long memory of volatility shocks exist in gold market volatility. Asymmetric models were again found to be superior when predicting stock index returns on Romania's BET index, where Gabriel (2012) found the asymmetric TGARCH specification to outperform the standard symmetric GARCH($1,1$) model in terms of forecast performance.

While GARCH models have long been the standard volatility prediction framework, new advances in data science and statistical methods paired with increased accessibility to high-performance computing, artificial neural networks (ANNs) have also been investigated and utilized for this purpose. ANNs have seen success in various fields including prediction, classification, and anomaly detection. The potential of ANNs for volatility prediction is highlighted in earlier works such as Donaldson and Kamstra (1997), where they demonstrated that an ANN could capture the asymmetric volatility effects of four large stock indexes around the world that were not captured by the GARCH models. Arnerić, et al. (2014) utilized a Jordan neural network (JNN), a type of recurrent neural network that is more complicated than a feedforward neural network but is known to excel in sequence prediction. Squared daily returns of the Croatian CROBEX index were used as inputs into the JNN model, as well a GARCH(*1,1*) model for comparison. The JNN model demonstrated superior performance in volatility prediction compared to the GARCH(*1,1*) model. Similar conclusions where ANNs have provided superior volatility predictions are seen when compared to implied volatility from S&P 500 Index futures options. Hamid and Iqbal (2004) found that the predicted volatility values from the ANN were not significantly different from the realized volatility, while the opposite was found for implied volatility as a prediction tool.

On the other hand, there are several examples where the ANN models have failed to outperform the GARCH-type models. Laily, et al. (2018) used an Elman Recurrent neural network (ERNN), which is functionally similar to the JNN discussed earlier. The authors found that the GARCH(*1,1*) model had a smaller mean squared error value than the ERNN model when forecasting volatility of stock prices. Likewise, Hossain, et al. (2009) also found the GARCH(*1,1*) model and the Support Vector Machine (SVM) machine learning model to both outperform a feedforward ANN in three of four markets. The authors also note that the GARCH and SVM models are more parsimonious than the ANN model, and require less data to accurately train.

Other works have created hybrid approaches to volatility prediction using hybrid GARCH-ANN models with promising results at the expense of parsimony. Kristjanpoller, et al. (2014) used an ANN to predict volatility on three South American stock exchanges, where the GARCH(*1,1*) output was used as an input into the ANN model. The hybrid ANN model realized a lower MAPE value when compared to the performance of the stand-alone GARCH model. Kristjanpoller and Hernández (2017) again applied this same GARCH-ANN hybrid approach to the metals markets, specifically prices of gold, silver, and copper. The study concluded again that the hybrid GARCH-ANN model improved forecasting performance when compared to the forecasts of the standalone GARCH models. When comparing the performance of hybrid GARCH-ANN models to each other, Lu, et al. (2016) found the EGARCH-ANN model to outperform other GARCH-ANN model hybrids.

This study builds upon this work in the current literature of comparing the volatility prediction performance of ANN models to GARCH models. However, the approach used in this study is different from the approaches in the literature as it focuses on identifying specific volatility profiles where the ANN or GARCH model outperforms the other. This study uses industry sector data as a proxy for unique volatility profiles.

# 3  Methodology

In order to determine the best fit GARCH and ANN models, we will estimate three GARCH specifications and three ANN architectures to determine which fit each sector's volatility profile best. The goal is to ensure the most appropriate specification or architecture is chosen for each model in order to identify if either model outperforms for a given sector.

## 3.1 GARCH Models

### 3.1.1 GARCH($p,q$) Model

The inability of linear models to explain features inherent to financial data such as volatility clustering, leverage effects, and leptokurtosis has forced the adoption of non-linear models to account for these features. The most common non-linear models used to forecast volatility in financial data are the ARCH family of models. The Autoregressive Conditional Heteroscedastic (ARCH) model is capable of modeling non-linear features such as the non-constant variances inherent in market time series data. The ARCH model is well suited for handling features such as volatility shocks, periods of volatility followed by continued volatility.

To develop the ARCH model mathematically, we begin with stock returns ($r_t$):

$$r_t = \mu + u_t \quad \text{where} \quad u_t = v_t \sigma_t \quad \text{and} \quad v_t \sim N(0,1) \qquad \text{\textit{Eq. 1}}$$

The returns in equations 1 are derived from the mean ($\mu$) plus a stochastic error term ($u_t$), where $v_t$ is a white-noise process following a normal distribution. The conditional variance is the variance of stock returns assumed to be dependant on the past stock return behaviour. The error term ($u_t$) can now be used to estimate the conditional variance (often referred to as conditional volatility) using the ARCH model in equation 2 (Engle, 1982).

$$\sigma_t^2 = a_0 + a_1 u_{t-1}^2 \qquad \text{\textit{Eq. 2}}$$

Equation 2 is referred to as an ARCH(1) model, where the conditional variance ($\sigma_t^2$) depends only on one squared lagged value of $u_t$ in equation 1. This ARCH(1) model can be modified into an ARCH($q$) model as in equation 3, where the conditional variance depends on $q$ number of lags (Brooks, 2014).

$$\sigma_t^2 = a_0 + a_1 u_{t-1}^2 + \cdots + a_q u_{t-q}^2 \qquad \text{\textit{Eq. 3}}$$

The ARCH model can quickly become unwieldy as one needs to determine the appropriate number of lags ($q$), which may result in a large number of lags and therefore a more complicated, less parsimonious model. The ARCH model also adheres to a non-negativity constraint, where all coefficients ($a$) are required to be non-negative. As the number of lags increase, it becomes more likely that this non-negativity constraint will be violated (Brooks, 2014).

A more parsimonious approach was developed by Bollerslev (1986), who built upon the ARCH model to develop a Generalized ARCH (GARCH) model. The GARCH model is similar to the ARCH model above, but the conditional variance is allowed to be dependant upon its own previous lags. Equation 4 formulates this approach:

$$\sigma_t^2 = a_0 + a_1 u_{t-1}^2 + \beta_1 \sigma_{t-1}^2 \qquad \text{\textit{Eq. 4}}$$

This is known as the GARCH($1,1$) model. This model adheres to the conditions $a_0 > 0$, $a_1 > 0$, $B_1 > 0$, and $a_1 + \beta_1 < 1$. The coefficients provide valuable insight into the behaviour of the conditional volatility. Coefficient $a_1$ indicates that volatility is sensitive to market shocks when the value is large (close to 1) and a large $\beta_1$ value indicates that the volatility persists and takes more time to die out.

The unconditional volatility ($\sigma^2$), or the long-term average variance, can also be calculated using the coefficients from equation 4. Conditional volatility will converge to the unconditional volatility over long timeframes. If the GARCH conditions above remain fulfilled, unconditional volatility can be calculated using equation 5.

$$\sigma^2 = \frac{a_0}{1 - (a_1 + \beta_1)}$$ *Eq. 5*

The coefficients from equation 4 can be used to measure the rate of convergence of conditional to unconditional volatility by simply summing $a_1 + \beta_1$.

Similar to the ARCH($q$) model, the GARCH model can be expanded to include more than one timestep. This extends the model to the GARCH($p, q$) version, which calculated conditional variance based on $q$ lags of the error variance and $p$ lags of the conditional variance term. This is identified in equation 6:

$$\sigma_t^2 = a_0 + \sum_{i=1}^{q} a_i u_{t-i}^2 + \sum_{j=1}^{p} \beta_j \sigma_{t-j}^2$$ *Eq. 6*

The GARCH($p,q$) model adheres to the same constraints above, but now applies to all $a_i$ and $\beta_j$ terms.

GARCH models require an alternative estimation method than those used for ordinary least squares (OLS) or other linear models. Since they are non-linear in form, the maximum likelihood (MLE) approach is used. MLE is a method that can identify the most likely values given a set of data, usually via finding the values that maximize a log-likelihood function. Once the mean equation (equation 1) and the appropriate GARCH model is identified (equation 4 or equation 6), then we can use the log-likelihood function under normality assumptions to find the appropriate parameter values in equation 7 (Brooks, 2014).

$$l = -\frac{T}{2}\ln(2\pi) - \frac{1}{2}\sum_{t=1}^{T} ln(\sigma_t^2) - \frac{1}{2}\sum_{t=1}^{T} \frac{u_t^2}{\sigma_t^2}$$ *Eq. 7*

GARCH models provide a solution to the issues of the ARCH model listed above. The GARCH model is more parsimonious, than that of its predecessor, and much less likely to breach the non-negativity constraints (Brooks, 2014). A GARCH(*1,1*) model for example, is equivalent to a ARCH($\infty$) model. With a GARCH(*1,1*) model, one does not need to estimate the number of lags and only needs to estimate a small number of parameters. The question often then arises of which GARCH model is appropriate to use. Brooks (2014) notes that a GARCH(*1,1*) model is sufficient to capture important features such as volatility clustering in financial data. This is backed up by the literature, where researchers often elect to use the GARCH(*1,1*) model in academics and industry. However, there are uses for higher order GARCH(*p,q*) models as well. Engle (2001) states that a higher order model (*p > 1 and/or q > 1*) is often more useful than a GARCH(*1,1*) model when using a long timeframe of data. In sum, it is important to test different specifications for each dataset to see which specification provides the most appropriate fit and accurately describes the volatility behaviour.

For the purposes of this study, all GARCH modelling is completed using the Python programming language and the ARCH library (Sheppard, et al., 2019).

### 3.1.2 Asymmetric GARCH Model

Volatility is assumed to follow a symmetric response to volatility shocks in ARCH models. This assumption is often violated in the real world, as a negative market shock during a time of crisis will usually cause volatility to rise more than a similar positive shock. This results in an asymmetric response to volatility shocks, usually caused by leverage effects as initially identified by Black (1976) or the volatility-feedback hypothesis. We can improve the GARCH specification and forecasts by accounting for volatility asymmetry if it is identified in the data.

There are many asymmetric GARCH models in use today. One of the most popular models is the exponential GARCH (EGARCH) model derived by Nelson (1991). The EGARCH conditional variance equation has several formulations, and this study will follow the formulation used in the Python ARCH library in equation 8 for an EGARCH(*p,q*) model:

$$\ln(\sigma_t^2) = \omega + \sum_{k=1}^{p} \beta_k \ln\left(\sigma_{t-k}^2\right) + \sum_{j=1}^{o} \gamma_j \frac{u_{t-j}}{\sigma_{t-j}} + \sum_{i=1}^{q} a_i \left[ \frac{|u_{t-i}|}{\sigma_{t-i}} - \sqrt{\frac{2}{\pi}} \right] \qquad \textit{Eq. 8}$$

The EGARCH model has the additional advantage of relaxing the non-negativity constraints imposed in the original GARCH models due to the inclusion of $\ln(\sigma_t^2)$. The indication of the level of asymmetry is also given via the gamma ($\gamma$) term. A significant gamma term indicates that asymmetry exists, and a negative gamma term indicates that the relationship between volatility and the returns are negative.

The EGARCH unconditional variance ($\sigma^2$) can be calculated by equation 9 (MathWorks, 2020).

$$\sigma^2 = \exp\left\{ \frac{\omega}{1 - \sum_{i=1}^{p} \beta_i} \right\} \qquad \textit{Eq. 9}$$

## 3.2   Artificial Neural Networks (ANN)

ANN models have seen success in the areas of market forecasting, equity trading, fraud detection, and more. By emulating the architecture of the neurons and their interconnections of the human brain, ANNs are renowned for their flexibility and capacity to find hidden patterns in noisy or incomplete data. This flexibility makes them a good candidate for forecasting volatility.

The ANN model used in this study is the Multi-Layer Perceptron (MLP). MLPs are widely used in industry and academics thanks to their simplicity. The MLP comprises of three main parameters: layers, neurons, and activation functions. The layers are made up of an input layer which reads in the data, a custom number of hidden layers which perform a deciding calculation, and an output layer that provides the forecasted output. Each layer is made up of a certain number of neurons, and the neurons utilize an activation function to make the calculations resulting in the predictions in the output layer. This is visualized in figures 2 and 3 for an arbitrary example network with one input layer with three neurons, one hidden layer with four neurons, and one output layer with one neuron to retrieve the final forecasted value.

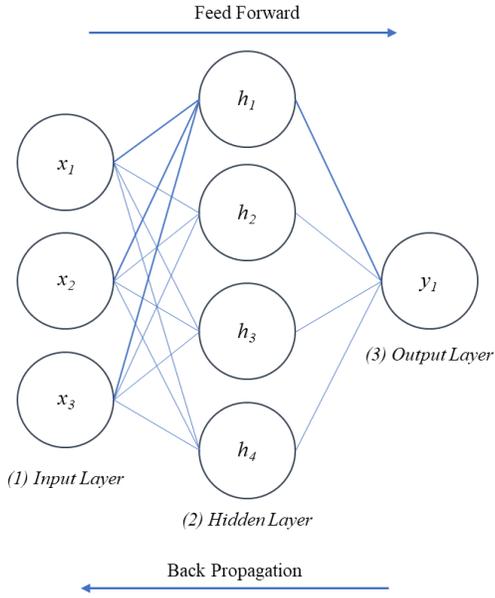

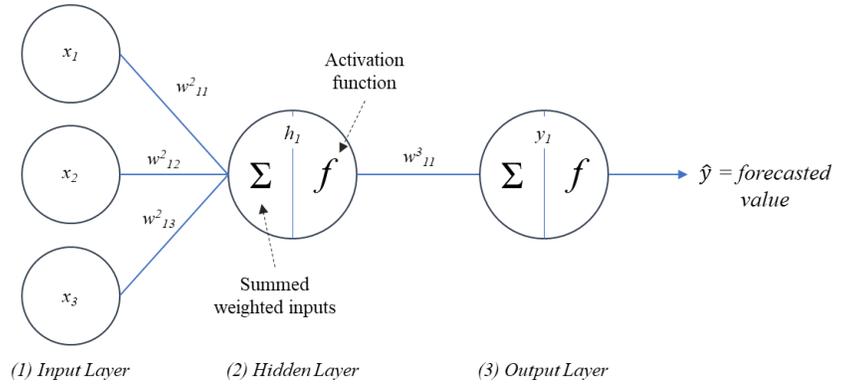

*Figure 2. A simplified illustration of a fully connected network with a three neuron input layer (x), a single hidden layer (h) with four neurons, and an output layer (y). The bolded lines correspond to the zoomed in figure 3.*

*Figure 3. A closer look at the relationship between the input layer (x) and a neuron of the hidden layer (h). This is simplified to a single hidden neuron to detail the calculations within.*

MLP networks are fully connected feedforward networks, as illustrated in figure 2. The name refers to how each neuron in each layer is connected to each neuron in the next layer. Figure 3 provides a closer look at the connections to the first neuron in the hidden layer from the input layer and illustrates how the data passes through the network to calculate a forecast. In this example, we will walk through the calculations for figure 3 where we have a single neuron in the hidden layer. The first stage of the neural network is the feedforward stage, where the input layer neurons ($x_p$) reads in the data and passes it through to the hidden layer. The input layer is made up of a vector of prior volatility values $p = (x_1, x_2, ..., x_p)$.

$$v_j = \sum_{i=1}^{p} w_{ji}^l x_i + b_j \qquad \text{Eq. 10}$$

The weight ($w$) has three parameters, where $j$ is the neuron in the $l^{\text{th}}$ layer from where the connection terminates, and $i$ is the neuron in the $(l-1)^{\text{th}}$ layer, where the connection originates (Nielsen, 2015). When multiple neurons are used in a layer, the weights ($w$) are represented in a matrix. The $b_j$ value represents a bias term and consists of a vector of constants for each layer that acts in a similar fashion to the intercept in a regression model. The initial bias term values and weights require no input from the user as they are initially generated randomly as small non-zero values. This results in the summed weights ($v$) feeding into the $j^{\text{th}}$ neuron.

Once the weights between the input and hidden layer have been calculated and summed, they pass to the activation function in the hidden layer. The activation function calculates the output of the neuron that will be fed into the next layer. For regression problems, the activation function outputs a forecasted value as opposed to a probability. The sigmoid activation function is one of the most common activation functions used in the existing literature and is also used in this study. The sigmoid function is a non-linear function that transforms the input $v_i$ from equation 10 into a value between 0.0 and 1.0 using equation 11. This is ideal, as the input volatility data has been scaled to 0.0 and 1.0 as well (see section 3.3.2).

$$h_q = f(v_j) = \frac{1}{1 + e^{-v_j}}$$

<div align="right"><em>Eq. 11</em></div>

The output from the sigmoid activation functions in the hidden layer creates a vector of outputs from each neuron $q = (h_1, h_2, \ldots, h_q)$. The process then moves to the next layer where we again calculate and sum the weights. The hidden layer output vector $q$ is the input into equation 10 again (replacing $x_i$ with $q_i$) and passing the resulting $v_j$ value to our sigmoid activation function in equation 11 to get our output scalar $\hat{y}$. This concludes the feedforward stage of the neural network.

The second stage is where the network learns and improves its forecast using the training data. The neural network moves to improve these predictions by using a loss function to compare the error between the predicted value and the forecasted value. The loss function used in this study is mean squared error (MSE), detailed in equation 13 where $y_i$ is the true volatility value at time $i$. MSE is the most used loss function for regression problems (Ding & Qin, 2020).

The error value is then distributed backwards through the network using the backpropagation algorithm to minimize the error between the true and predicted values. Originally popularized by Rumelhart, Hinton, and Williams (1986), the backpropagation method has become the workhorse of most neural network architectures today. Since the weights of the ANN are the only variable elements that can be changed in order to influence the next neurons and the ultimately the predicted value, this algorithm is an efficient optimization method. Backpropagation uses the chain rule to find the rate of change, or gradient, of the loss function with respect to each weight. Each weight is then optimized layer by layer as the algorithm propagates backwards from the output node to the input layer. Once all weights have been updated, the process starts a new feedforward pass with the new weights in place, and the output is compared to the true values once again. This process repeats until each weight settles to a value that results in the lowest MSE between the true and predicted values. Each forward and back pass through the network is referred to as an epoch, where a large number of epochs give the network more time to learn and adjust weights to minimize the loss function.

ANNs provide an alternative method to forecast volatility than GARCH models. The ANNs advantage lies in its ability to model complex non-linear relationships while not imposing any restrictions on the data or the model variables. These advantages make ANNs well suited for predicting values from data with heteroscedastic or autocorrelated features, such as volatility. All ANN modelling in this study is completed using the Python programming language and the TensorFlow Deep Learning library (Abadi, et al., 2019).

## 3.3 Estimation and Prediction Methodology

### 3.3.1 GARCH Estimation and Prediction

To retrieve the most accurate predictions from the GARCH models, we estimate three GARCH models and select the model with the best fit as the model to proceed with predictions for each sector. A GARCH(*1,1*) model is estimated for each sector, as this model is often regarded in the literature as the most useful GARCH specification for its simplicity and ability to fit a wide range of data. However, this model is not always the most appropriate, so a custom algorithm is used to also estimate a higher-order GARCH(*p,q*) and EGARCH(*p,q*) model (where *p >1 and/or q >1*). This algorithm solves for the combination of *p* and *q* with the lowest AIC and white noise residuals. This specification allows better flexibility to capture the heteroscedasticity and autocorrelation effects in the data at the risk of a more complicated model. Lastly, an EGARCH model will be estimated to evaluate if the model needs to account for asymmetry in the data.

The evaluation of each model will be based on how well the heteroscedasticity and autocorrelation (referred to as ARCH effects) are captured in the data, provided no constraints are violated. The use of statistical tests such as the Ljung-Box Q statistic will provide this information. Forecasting ability will be measured by the Akaike Information Criteria (AIC). The AIC allows for a direct comparison between GARCH specifications, and has been shown to work well with GARCH(*1,1*) as well as higher-order models (Naik, et al., 2020).

Once the appropriate specification has been determined, a rolling window forecast will be used to predict the conditional volatility value one step ahead ($\sigma_{t+1}^2$). The rolling window forecast initially uses a fixed sample length, then forecasts the value one step ahead from the final observation. The next forecast then drops the first value of the fixed sample and includes the true value of the prior forecasted value as the final value, and so on.

### 3.3.2 ANN Estimation and Prediction

The example network given in section 3.2 differs slightly from the ANNs used in this study. The ANN architecture used in this study has an input layer with five neurons, as we will be inputting a vector of prior volatility values from a five day look back period. Several lookback windows were examined, with the five-day lookback period producing the best results. The hidden layer has a varying number of neurons, and the output layer always has a single neuron where we retrieve the forecasted value.

The ANN estimation and prediction follows the same process as the GARCH model, where we choose the best fit ANN model from three different architectures. The architectures differ in the number of neurons in the hidden layer, ranging from 1, 12, or 50 neurons. Choosing the correct number of neurons is a difficult task as there is less consensus on a standard procedure to do so. ANN users usually resort to trial and error and experimentation to find the optimal number of neurons, so that is what we have done here.

The prediction process begins with the data being read into an auto-scaler that transforms each prior volatility data point to a range between 0.0 and 1.0. This preprocessing step is important as it speeds up the training process and allows for more robust weights that will be learned when training the network, further stabilizing the model. The vector of transformed prior volatility values is then fed into the input layer, where the process in section 3.2 takes over. The ANN models use the vector of five prior volatility values to predict the next single volatility value ($\sigma_{t+1}^2$) one step ahead. This is the same process as outlined for GARCH prediction in section 3.3.1, where the same rolling window method is used.

## 3.4 Prediction Performance Metrics

This study uses the three common performance metrics of Mean Absolute Error (MAE), Mean Squared Error (MSE), and Root Mean Squared Error (RMSE). For all metrics, the lower the resulting value the better.

MAE calculates the absolute distance between the forecasted value and the true value. MAE has the advantage in situations where sensitivity to outliers is not of concern since it does not alter the error values. The errors calculated from MAE can be directly compared to the target variable as they are in the same units. MAE is calculated using equation 12, where $y$ is the true value and $\hat{y}$ is the predicted value.

$$MAE = \frac{1}{N}\sum_{i=1}^{N}|y_i - \hat{y}_i|$$

*Eq. 12*

MSE and RMSE calculate the error differently giving the metric unique characteristics compared to MAE. Under MSE, the distance between the true value and the forecasted value is squared. This results in much more weight being attached to larger errors, and less weight attached to smaller errors. In general, MSE measures the average squared difference of the true value and forecasted value, resulting in a positive metric. The RMSE is the square root of the MSE and has the advantage of transforming the MSE metric into the same unit as the target variable, making them directly comparable. MSE and RMSE are calculated using equations 13 and 14 respectively.

$$MSE = \frac{1}{N}\sum_{i=1}^{N}(y_i - \hat{y}_i)^2$$

*Eq. 13*

$$RMSE = \sqrt{\frac{1}{N}\sum_{i=1}^{N}(y_i - \hat{y}_i)^2}$$

*Eq. 14*

RMSE will serve as the main measure to compare GARCH and ANN volatility prediction performance in this study as it is generally the preferred performance metric for regression-type problems (Géron, 2019). The RMSE is ideal in this study since it gives a higher weight to larger errors. This is important in volatility prediction as large shocks in volatility often take time to dissipate, so it is very important to accurately capture these shocks in predictions. Large errors in this case are more undesirable and so warrant the extra weighting by RMSE.

# 4 Dataset Overview

A proxy for volatility profiles is required to be used for prediction performance comparisons. This study uses industry sectors with unique volatility characteristics for this purpose. The five industry portfolio dataset from the Kenneth R. French data library provides the sector return data (French, 2020). The dataset provides daily returns and is comprehensive in containing all stocks in the NYSE, AMEX, and NASDAQ exchanges. It is sorted into five broad sectors based on the Standard Industrial Classification (SIC) codes, which are consumer durables, health, technology, manufacturing, and other.

The consumer durables sector portfolio contains all stocks in wholesale, retail, and consumer services. The health sector portfolio mainly contains all medical equipment firms and drug companies. Technology contains all business equipment, phone, television transmission, and information processing companies. The manufacturing sector holds all manufacturing, energy, and utility stocks. All other companies are categorized into the other sector, which contains finance, entertainment, mines, construction, business services, transportation, and construction.

## 4.1   Sector Statistics

The total dataset contains the daily arithmetic returns for each sector, spanning over 15 years from January 3rd, 2005 to April 30th, 2020. The total sample size contains 3858 daily observations. Important volatility time periods are included, such as the 2008 market crash, 2011 correction, and most recently the 2020 volatility during the Covid-19 global pandemic and oil market disruptions. Daily observations are chosen as they are commonly used in industry and academics to capture day-to-day changes in the markets.

The returns of each sector are calculated by French using equation 15:

$$R_t = \frac{P_t - P_{t-1}}{P_{t-1}} \qquad\qquad Eq.\ 15$$

For the purposes of model specification and forecasting, the data is split into an in-sample training set and an out-of-sample test set. The in-sample training set contains 80% of the dataset to be used for model training, 3086 observations in total. The out-of-sample test set contains the remaining 20% to be used for forecasting, 772 observations.

The sector in-sample and out-of-sample returns are presented in figure 4. Timeframes of extreme market turbulence are clearly visible in each chart. This includes notable periods such as the 2008 financial crises, the market correction in 2011-2012, the oil price crash in 2016, and the most recent period of high uncertainty due to the Covid-19 pandemic alongside another oil price crash in early 2020. The different levels of volatility in each sector can also be distinguished. For instance, the 2008 financial crises is much more pronounced in the other sector, compared to the consumer durables sector. This is mainly due to financial stocks being included in the other sector, and those stocks in being highly affected by the crisis. It is apparent that large dispersions in returns during turbulent times in the market are often followed by further large dispersions in returns. This behaviour may be indicative of ARCH effects.

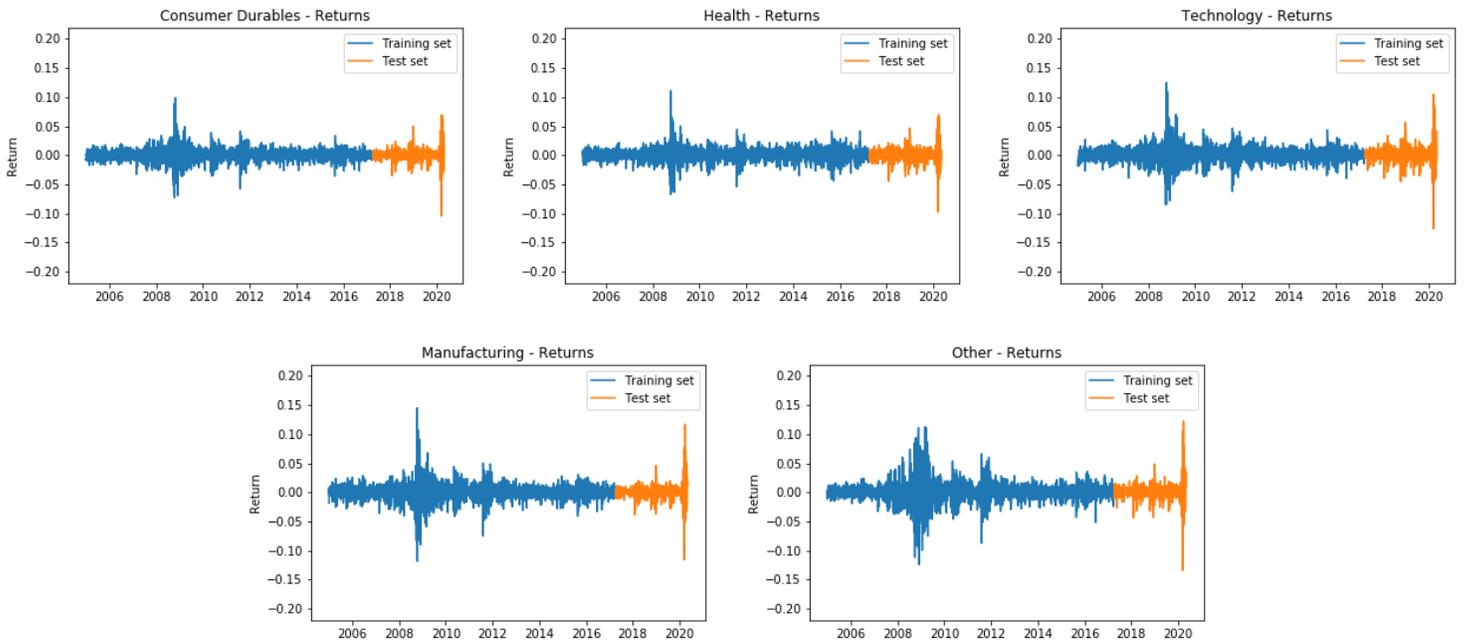

*Figure 4. Sector returns*

These indications are further investigated in table 1 which provides descriptive statistics for the in-sample dataset. The mean across all three sectors is close to zero, while the standard deviation gives rough insight into their unconditional volatility. This is also visible through the increasing min/max spread of the sectors as the standard deviation becomes larger. Skewness and kurtosis indicate non-normality, which is confirmed by the Jarque-Bera (JB) statistic for each sector, rejecting the null hypothesis of normality at the 1% level. The Ljung-Box Q statistic rejects the null hypothesis that the data is independently distributed up to the 12[th] lag for all sectors, indicating the presence of autocorrelation. The earlier indication of ARCH effects is further reinforced by the Ljung-Box Q$^2$ statistic, where the squared returns indicate the presence of heteroscedasticity with 12 lags. The Augmented Dickey Fuller (ADF) test rejects the null hypothesis of a unit root in all sectors at the 1% level, indicating the returns are stationary.

| Descriptive Statistics | | | | | | Statistical Tests | | | |
|---|---|---|---|---|---|---|---|---|---|
| | Mean | Std. Dev | Skewness | Kurtosis | Max | Min | J-B | Q (12) | Q$^2$ (12) | ADF |
| Consumer Durables | 0.00043 | 0.01037 | -0.02916 | 8.82223 | 0.09880 | -0.07270 | 9963.6** | 49.21** | 2937.2** | -12.73** |
| Health | 0.00043 | 0.01057 | -0.04208 | 8.00669 | 0.11100 | -0.06700 | 8206.1** | 55.56** | 2016.0** | -14.64** |
| Technology | 0.00044 | 0.01261 | 0.12667 | 9.41801 | 0.12470 | -0.08510 | 11362.3** | 41.644** | 3029.5** | -12.20** |
| Manufacturing | 0.00042 | 0.01366 | -0.07920 | 12.54426 | 0.14520 | -0.11860 | 20151.6** | 66.74** | 3577.2** | -13.00** |
| Other | 0.00031 | 0.01609 | -0.05635 | 9.90497 | 0.11270 | -0.12430 | 12562.4** | 60.43** | 3355.5** | -11.01** |

*Table 1 - ** significant at the 1% level*
*\* significant at the 5% level*

The sectors can be sorted into a broad categorization of high, medium, and low volatility based on their standard deviation in table 1. The low volatility sectors are consumer durables and health, while technology corresponds to the medium volatility category as its standard deviation is very close the average of all five sectors, while the manufacturing and other sector fit into the high volatility category. The true long run average volatility of each sector is further examined in section 5.1 using the unconditional volatility calculations.

## 4.2    Volatility Proxy

Volatility is often interpreted as risk or uncertainty, and as such is directly unobservable in the market. A proxy must be used to provide a value for the true volatility to which the model predictions will be compared. The most common method of estimating volatility is by calculating the variance of the returns as given by equation 16.

$$\sigma^2 = \frac{\sum_{t=1}^{n}(r_t - \bar{r})^2}{n-1} \ \ where \ \ \bar{r} = \frac{\sum_{t=1}^{n} r_t}{n} \qquad\qquad Eq.\ 16$$

However, the most common volatility proxy for forecasting and modelling is simply the squared returns (Brooks, 2014). This approach essentially assumes a mean return ($\bar{r}$) of zero and provides a volatility proxy estimate for each specific day. Andersen and Bollerslev (1998) elect for using squared intra-day return over squared daily return since the intra-day calculation ignores the overnight stock movements and only includes the differences from market open to market close. Since the intra-day trading data is not available for the selected portfolios, the squared daily returns will be used as the true volatility proxy as outlined in equation 17. This approach is also parsimonious and widely used in the literature.

$$\sigma^2 = r_t^2 \qquad\qquad Eq.\ 17$$

# 5 Results

## 5.1 GARCH Estimation: In-Sample Data

For each sector, the three GARCH specifications outlined in section 3.1 are estimated, namely the GARCH($p,q$), GARCH($1,1$), and EGARCH($p,q$) models. The best specification is chosen for forecasting based on the adequacy of fit and AIC criterion. Each GARCH model is estimated using the in-sample portion of the daily return data to prevent look-ahead bias in the estimations. Results for all sectors are displayed in Table 2.

Of the three GARCH models estimated, the EGARCH($p,q$) model is the most adequate model to proceed with forecasting for all five sectors. These EGARCH specifications are highlighted in table 2. Each EGARCH specification results in a conditional mean equation that is not significantly different from zero at least at the 1% level, while the conditional volatility equations result in varying numbers of significant coefficients. The EGARCH models adequately capture all ARCH effects in the returns as given by the insignificant Q-statistics at least at the 1% level. This indicates that the resulting residuals and squared residuals are a white noise process. Although the EGARCH model relaxes the non-negativity constraints, the $\beta$ terms must be positive and less than one to remain stationary, which is the case in all EGARCH models (Ezzat, 2012).

The main benefit of the EGARCH model is the ability to account for potential asymmetries in the response to volatility shocks via the gamma term. Each EGARCH model has a negative gamma term that is significantly different from zero at the 1% level, which implies that volatility asymmetry exists and the relationship between volatility and returns is negative. This relationship indicates that leverage effects exist in the data of each sector, where a negative shock has a greater effect on volatility than a similar positive shock (Brooks, 2014). Once asymmetry have been identified, it is important to account for it when forecasting future volatility values. The forecasting potential for each model is also given by the AIC value, where the EGARCH models provide a lower value than the other GARCH models across all sectors. This indicates the EGARCH specification have the best forecasting ability for this study.

The GARCH($p,q$) and GARCH($1,1$) model estimations for each sector resulted in varying levels of model adequacy. The GARCH($p,q$) models, despite having differing $p$ and $q$ values, and GARCH($1,1$) models resulted in very similar coefficients for all sectors except other. This shows that the unconditional volatility of each sector when estimated by these specifications is very similar across four sectors. Both models satisfy the non-negativity constraints imposed by the standard GARCH models, and all have coefficients that sum to less than one. The large $\beta$ coefficients in these GARCH models of each sector indicate that volatility persists for a long time after a shock. Similarly, since the coefficients sum to a number close to one, the conditional volatility reverts to the unconditional (long term average) volatility very slowly. However, in all sectors except Other, the low $\alpha$ coefficients indicate that the initial reaction of conditional volatility to market shocks is quite minimal in the first place.

Despite the insights the GARCH($p,q$) and GARCH($1,1$) models provide into the volatility characteristics of the sectors, they are not as adequate as the EGARCH specifications for forecasting. The GARCH($p,q$) and GARCH($1,1$) models fail to capture ARCH effects in several sectors as indicated by the significant Q statistics at the 1% level, while both models have much higher AIC values than the EGARCH specifications.

### Consumer Durables

|  | GARCH(2,1) | GARCH(1,1) | EGARCH(3,5) |
|---|---|---|---|
| $\mu$ | 0.000596** | 0.000609 | 0.000209 |
| $\alpha_0$ | 0.000002** | 0.000002 | -0.3560** |
| $\alpha_1$ | 0.0500* | 0.1000 | -0.00158 |
| $\alpha_2$ | 0.0500 |  | 0.1195* |
| $\alpha_3$ |  |  | 0.0722 |
| $\alpha_4$ |  |  |  |
| $\gamma$ | NA | NA | -0.1978** |
| $\beta_1$ | 0.8800** | 0.8800 | 0.8234** |
| $\beta_2$ |  |  | 0.000000 |
| $\beta_3$ |  |  | 0.000000 |
| $\beta_4$ |  |  | 0.000000 |
| $\beta_5$ |  |  | 0.1386 |
| Coeff. Sum: | 0.98 | 0.98 | 0.60 |
| Q (12) | 24.00* | 23.86* | 24.99* |
| Q2 (12) | 16.27 | 19.32 | 8.98 |
| AIC | -20674.50 | -20672.9 | -20791.0 |

### Health

|  | GARCH(2,1) | GARCH(1,1) | EGARCH(2,1) |
|---|---|---|---|
| $\mu$ | 0.000639** | 0.000650** | 0.000286 |
| $\alpha_0$ | 0.000002** | 0.000002** | -0.3563** |
| $\alpha_1$ | 0.0500** | 0.1000** | 0.0142 |
| $\alpha_2$ | 0.0500** |  | 0.1338** |
| $\alpha_3$ |  |  |  |
| $\alpha_4$ |  |  |  |
| $\gamma$ | NA | NA | -0.1396** |
| $\beta_1$ | 0.8800** | 0.8800** | 0.9617** |
| Coeff. Sum: | 0.98 | 0.98 | 0.61 |
| Q (12) | 19.54 | 19.69 | 22.64* |
| Q2 (12) | 22.15* | 28.37** | 17.73 |
| AIC | -20258.9 | -20254.3 | -20360.6 |

### Technology

|  | GARCH(2,1) | GARCH(1,1) | EGARCH(4,3) |
|---|---|---|---|
| $\mu$ | 0.000827** | 0.000780** | 0.000332* |
| $\alpha_0$ | 0.000003** | 0.000003** | -0.4051** |
| $\alpha_1$ | 0.0500* | 0.1000** | 0.006889 |
| $\alpha_2$ | 0.0500 |  | 0.0949 |
| $\alpha_3$ |  |  | 0.0452 |
| $\alpha_4$ |  |  | 0.0591 |
| $\gamma$ | NA | NA | -0.2222** |
| $\beta_1$ | 0.8800** | 0.8800** | 0.6851** |
| $\beta_2$ |  |  | 0.000000 |
| $\beta_3$ |  |  | 0.2702* |
| Coeff. Sum: | 0.98 | 0.98 | 0.53 |
| Q (12) | 16.8 | 16.99 | 16.89 |
| Q2 (12) | 12.85 | 15.55 | 7.94 |
| AIC | -19562.5 | -19560.8 | -19693.5 |

### Manufacturing

|  | GARCH(2,1) | GARCH(1,1) | EGARCH(2,5) |
|---|---|---|---|
| $\mu$ | 0.000643** | 0.000657** | 0.000258 |
| $\alpha_0$ | 0.000004 | 0.000004** | -0.2384** |
| $\alpha_1$ | 0.0500 | 0.1000** | 0.0282 |
| $\alpha_2$ | 0.0500 |  | 0.1492** |
| $\alpha_3$ |  |  |  |
| $\alpha_4$ |  |  |  |
| $\gamma$ | NA | NA | -0.1513** |
| $\beta_1$ | 0.8800** | 0.8800** | 0.8316** |
| $\beta_2$ |  |  | 0.000000 |
| $\beta_3$ |  |  | 0.000000 |
| $\beta_4$ |  |  | 0.000000 |
| $\beta_5$ |  |  | 0.1417 |
| Coeff. Sum: | 0.98 | 0.98 | 0.76 |
| Q (12) | 13.9 | 14.30 | 17.36 |
| Q2 (12) | 21.03* | 27.21** | 11.18 |
| AIC | -19254 | -19249.8 | -19336.3 |

### Other

|  | GARCH(4,1) | GARCH(1,1) | EGARCH(3,1) |
|---|---|---|---|
| $\mu$ | 0.000675** | 0.000749** | 0.000329* |
| $\alpha_0$ | 0.000003** | 0.000005** | -0.1126** |
| $\alpha_1$ | 0.0695* | 0.2000** | 0.1036** |
| $\alpha_2$ | 0.0506 |  | 0.1137 |
| $\alpha_3$ | 0.000001 |  | -0.0799 |
| $\alpha_4$ | 0.000001 |  |  |
| $\gamma$ | NA | NA | -0.1046** |
| $\beta_1$ | 0.8651** | 0.7800** | 0.9870** |
| Coeff. Sum: | 0.99 | 0.98 | 0.91 |
| Q (12) | 22.29* | 21.00 | 20.84 |
| Q2 (12) | 10.82 | 21.69* | 15.93 |
| AIC | -18843.8 | -18798.8 | -18916.1 |

*Table 2. Results from the GARCH estimations of each specification for all sectors. Adequacy of each model is determined by the coefficients, Q-statistics, and AIC criteria.*

*** significant at the 1% level*
* *significant at the 5% level*

The broad categorization of the sectors into high, medium, and low volatility discussed in section 4.2 can now be expanded upon since earlier we were only able to use standard deviation as a volatility indicator. Since the most appropriate GARCH model for each sector has been identified, we can calculate the unconditional variances. This provides a better overall volatility measure since it represents the long run average volatility. The unconditional volatility of each sector's EGARCH specifications is calculated using equation 9. The results are presented in table 3.

| | Sector | Unconditional Variance ($\sigma^2$) | Unconditional Std. Dev. ($\sigma$) |
|---|---|---|---|
| Low Volatility | Consumer Durables | 0.0000854 | 0.0092400 |
| | Health | 0.0001077 | 0.0103792 |
| Medium Volatility | Technology | 0.0001159 | 0.0107665 |
| High Volatility | Manufacturing | 0.0001325 | 0.0115114 |
| | Other | 0.0001731 | 0.0131574 |

*Table 3. The unconditional variance and standard deviation (volatility) calculated from the EGARCH estimation of each sector.*

Our initial categorization attempt based only on the sector's standard deviation resulted in the same categorization as the unconditional volatility measure in table 3. This categorization will help determine the performance of GARCH and ANN predictions within these broader three volatility profiles and initial five sector categories.

## 5.2   ANN Estimation: In-Sample Data

Similar to the GARCH model specification, three neural network architectures are trained on the in-sample training dataset. The same train-test split is used as in the GARCH models identified in section 4.2, except 10% of the training data is used as the validation set. The neural network specification process does not have clear guidelines as in the case of GARCH models. The methods used in ANN model specification is a combination of following the current literature in ANN architecture and a fair bit of trial-and-error experimentation. The multi-layer perceptron (MLP) used in this study tests the performance of three ANN architectures with one input, hidden, and output layer, but with a varying number of neurons.

Two main decisions are required to build a neural network, namely the number of hidden layers and the number of neurons contained therein. A single hidden layer is often sufficient in practice and is widely used in the literature. One hidden layer has the benefit of simplicity while still maintaining the ability to approximate complex functions (Heaton, 2008). A single hidden layer also provides an indicator of an ANNs practicality for a particular problem. If a model performs well with a simple single hidden layer network, this indicates further refinement or use of more complex ANNs may improve performance further. For the reasons above we will be using a single hidden layer in this study. There is less consensus on the methods to calculate the appropriate number of neurons in each hidden layer, and this is usually done by experimentation. As such, we will be testing ANNs with 1, 12, and 50 neurons in the hidden layer to identify the best performing ANN. In each architecture, there is always five nodes in the input layer and a single node in the output layer as identified in section 3.3.2.

The in-sample training results are detailed in table 4. The performance of each model is gauged by the MAE, MSE, and RMSE metrics. As explained in section 3.3, the lowest RMSE value will ultimately decide which model is the best specification for that sector and move forward with volatility predictions. A relatively low number of 60 epochs are used to reduce the chance of overfitting the model as well as to relieve the computational burden of training the network.

| *In-Sample Train Set* | | **MAE** | **MSE** | **RMSE** |
|---|---|---|---|---|
| | **(5,1,1)** | 0.0001262 | 0.0000001 | 0.0003323 |
| **Consumer Durables** | **(5,12,1)** | 0.0001267 | 0.0000001 | 0.0003316 |
| | **(5,50,1)** | 0.0001683 | 0.0000001 | 0.0003498 |
| | **(5,1,1)** | 0.0001250 | 0.0000001 | 0.0003367 |
| **Health** | **(5,12,1)** | 0.0001321 | 0.0000001 | 0.0003328 |
| | **(5,50,1)** | 0.0001112 | 0.0000001 | 0.0003441 |
| | **(5,1,1)** | 0.0001914 | 0.0000003 | 0.0005126 |
| **Technology** | **(5,12,1)** | 0.0001854 | 0.0000003 | 0.0005123 |
| | **(5,50,1)** | 0.0001815 | 0.0000003 | 0.0005232 |
| | **(5,1,1)** | 0.0002330 | 0.0000004 | 0.0006628 |
| **Manufacturing** | **(5,12,1)** | 0.0002115 | 0.0000004 | 0.0006465 |
| | **(5,50,1)** | 0.0001985 | 0.0000005 | 0.0006734 |
| | **(5,1,1)** | 0.0003326 | 0.0000007 | 0.0008428 |
| **Other** | **(5,12,1)** | 0.0003121 | 0.0000007 | 0.0008421 |
| | **(5,50,1)** | 0.0003045 | 0.0000007 | 0.0008614 |

*Table 4 – In-sample neural network training results to choose the best specification for forecasting.*

In terms of RMSE, the model with 12 neurons (denoted NN(5,12,1)) outperformed all other architectures in each sector. It is interesting to note that the training error appears to improve when the number of neurons is increased to 12, but then appears to worsen when a higher number of neurons is used, as in the case of 50 neurons. This could be due to the model beginning to overfit the data as the number of neurons increases. The 12 neuron architecture is highlighted as the best specified model for volatility prediction for each sector.

We can now check to ensure that these models do not overfit the training data. This will help ensure the chosen models provide robust forecasts as new out-of-sample data is introduced. Figure 5 displays the learning curves for the training and validation data for the NN(5,12,1) models. The learning curves for each sector show training loss is continually decreasing then reaching a stable continuous value where the line begins to flatten, while the validation curve has similar behaviour. This indicates the models are not overfit and are generalizing the data well, not simply memorizing the patterns in the training data. The generalization gap between the training loss and validation loss is due to the validation data not being span over periods of particularly high volatility shocks, and so produces a lower error than the training set which includes those time periods. However, the gap between the training and validation loss remains quite small.

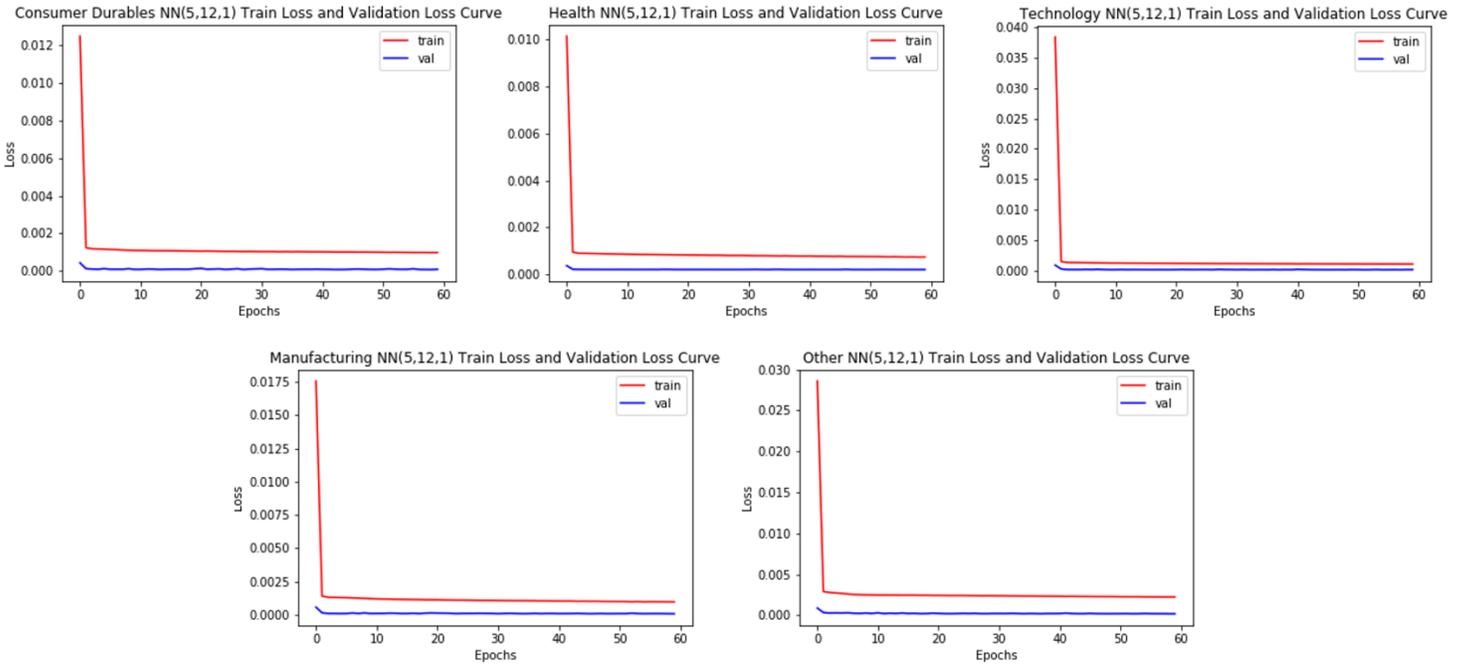

*Figure 5. NN(5,12,1) learning curves*

## 5.3 GARCH and ANN Forecast Comparison: Out-of-Sample Data

Following the GARCH and ANN specification process, the chosen models for each sector are used to predict the future sector volatility values of the out-of-sample test dataset. The test dataset contains 20% of the overall values from April 2017 to April 2020, including the unusually high volatility periods in early 2020. The results of the sector volatility predictions are presented in table 5 with the best performing model type highlighted for each sector.

The performance between the GARCH and ANN models is very comparable in terms of the MAE, MSE, and RMSE metrics. As MAE does not place any additional weight to outliers, it provides a much lower value than RMSE. As detailed in section 3.3, this study uses RMSE as the main performance decision metric as it assigns extra weight to outliers such as the extremely high volatility periods.

The GARCH models provide a lower MAE value for each sector indicating the GARCH model predictions were on average closer to the true values in absolute terms compared to the ANN model. RMSE, the main evaluation metric for this study, on the other hand shows that the GARCH model performs better only in the technology, manufacturing, and other sectors. The RMSE results for the consumer durables and health sectors indicate that an ANN model provides a better prediction with this volatility profile.

| Consumer Durables | MAE | MSE | RMSE |
|---|---|---|---|
| EGARCH(3,5) | 0.0001291 | 0.0000003 | 0.0005327 |
| ANN(5,12,1) | 0.0001530 | 0.0000003 | 0.0005258 |

| Health | MAE | MSE | RMSE |
|---|---|---|---|
| EGARCH(2,1) | 0.0001479 | 0.0000002 | 0.0004747 |
| ANN(5,12,1) | 0.0001720 | 0.0000002 | 0.0004655 |

| Technology | MAE | MSE | RMSE |
|---|---|---|---|
| EGARCH(4,3) | 0.0002108 | 0.0000006 | 0.0007981 |
| ANN(5,12,1) | 0.0002447 | 0.0000006 | 0.0008038 |

| Manufacturing | MAE | MSE | RMSE |
|---|---|---|---|
| EGARCH(2,5) | 0.0001788 | 0.0000006 | 0.0007858 |
| ANN(5,12,1) | 0.0002285 | 0.0000006 | 0.0008047 |

| Other | MAE | MSE | RMSE |
|---|---|---|---|
| EGARCH(3,1) | 0.0002231 | 0.0000008 | 0.0009217 |
| ANN(5,12,1) | 0.0002734 | 0.0000009 | 0.0009698 |

*Table 5 – Out-of-sample prediction performance results between the GARCH and ANN models that were specified for each sector in sections 5.1 and 5.2.*

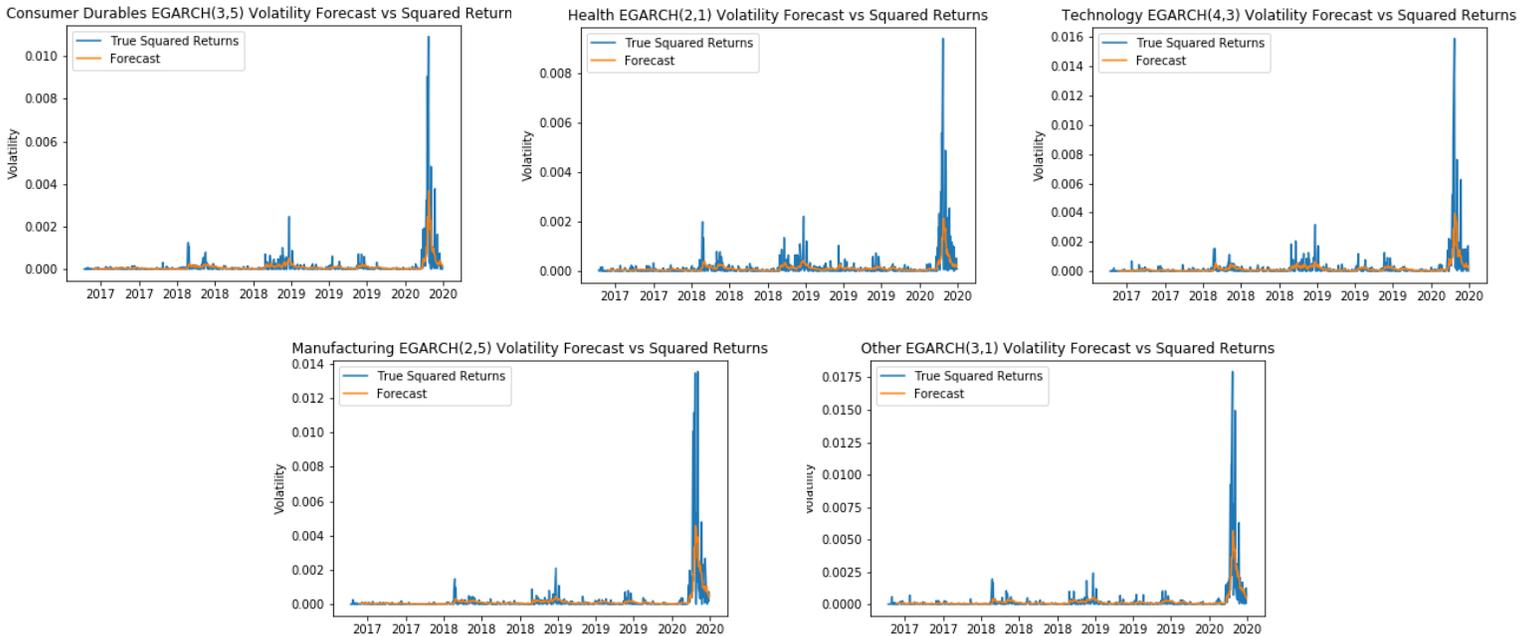

*Figure 6. GARCH predictions plotted against the true volatility values.*

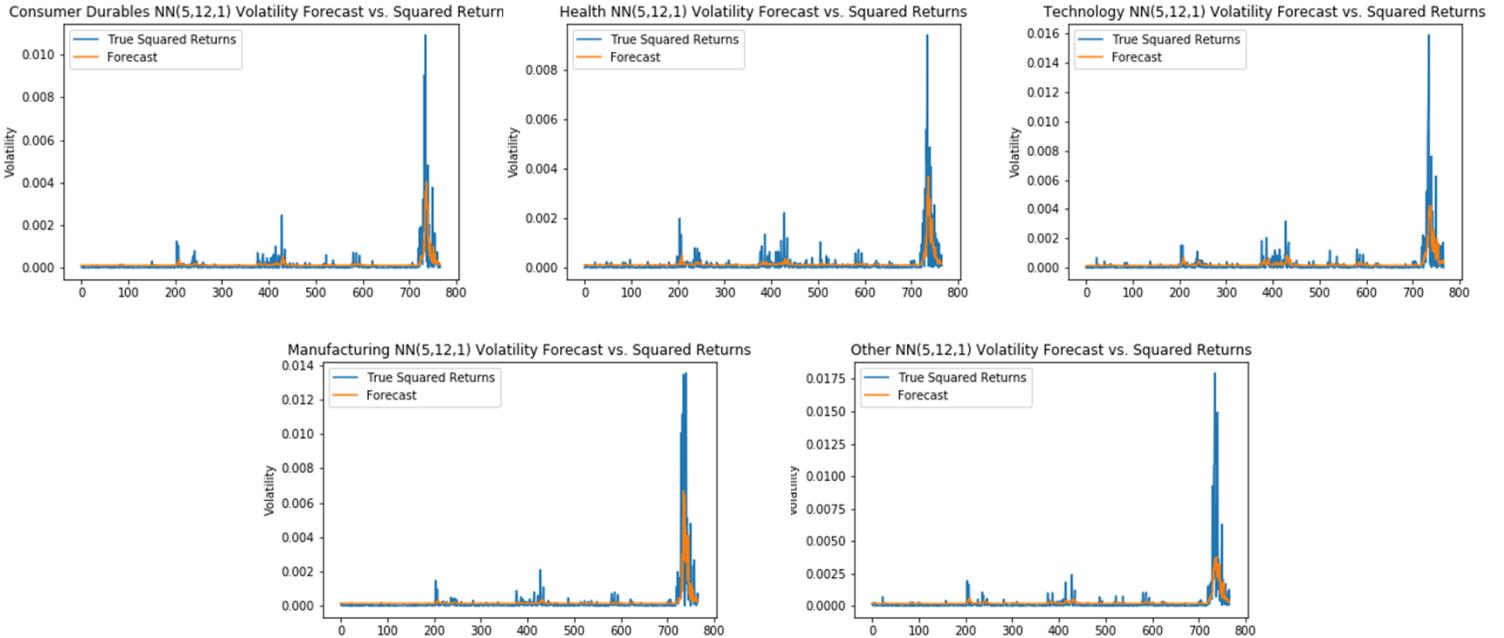

*Figure 7. ANN predictions plotted against the true volatility values.*

Figures 6 and 7 plot the forecasted values against the actual volatility proxy for the GARCH and ANN models respectively. Both the GARCH and ANN predictions closely track the true squared return values. The orange forecast line follows the volatility spikes quite closely in all predictions, with the most prominent spike following the volatility spike in early 2020. The ANN model appears to forecast larger values, most notably visualized when comparing the large volatility spike in the manufacturing and health sector. The ANN predicted values reach higher than the GARCH predicted values as indicated by a taller orange forecast line. Both GARCH and ANN volatility predictions appear to lag behind the actual squared returns by a few days.

# 6  Discussion of Results

The initial GARCH modelling of three different GARCH specifications (GARCH(*1,1*), GARCH(*p,q*), EGARCH(*p,q*)) shows the EGARCH model providing the most adequate fit for all sectors. The main benefit of the EGARCH model is its ability to account for asymmetry in the return data, where the returns in each sector are found to have significant asymmetry. By accounting for this, the EGARCH specifications are shown to have the best forecasting potential as per the AIC value and are chosen for forecasting. This is in line with previous studies such as Lin (2018), Miron and Tudor (2010), and Chong, et al. (1999) where the EGARCH model was shown to provide a better fit to the data and better forecasting performance for stock market return volatility than other GARCH models.

The ANN modelling followed a similar process, where the architecture with 12 neurons in the hidden layer is shown to outperform the architecture with one or 50 neurons. The results of fitting in the in-sample data in table 4 show that the models with one neuron in the hidden layer provided slightly larger errors than the model with 12 neurons , while the models with 50 neurons provided much larger errors. This indicates that in this case, more neurons do not increase model accuracy. It is known that a larger number of neurons increases the model complexity and the risk of overfitting the data (Goodfellow, et al., 2016). This increase in model complexity may increase performance of ANNs in some applications, but this result indicates that increased model complexity decreases performance when predicting volatility. Given these results, we can be confident that the ideal of parsimony applies here, where a simpler 12 neuron model has outperformed more complicated models.

The resulting out-of-sample forecasting performance of the EGARCH($p$,$q$) specifications and the ANN(5,12,1) models is compared in table 5. As discussed in section 3.4, the RMSE measure is chosen as the deciding metric to evaluate model prediction performance in this study. The results show the ANN model provides more accurate volatility predictions in the consumer durables and health sectors, while the GARCH model outperforms the ANN model in the technology, manufacturing, and other sectors. The classification of these sectors into high, medium, and low volatility profiles in sections 4.1 and 5.1 reveals that the ANN model outperforms the GARCH model when predicting volatility in low volatility sectors (consumer durables and health), while the GARCH model outperforms the ANN model in the medium (technology) and high volatility sectors (manufacturing and other).

The RMSE performance of both models is very close in the medium volatility sector. However, the ANN underperformed the GARCH model in the high volatility sectors by a much larger RMSE margin. As visualized in figure 4, the high volatility sectors have larger volatility shocks (manufacturing) or the shocks take longer to dissipate (other). This indicates that the EGARCH specifications are better at forecasting volatility over time periods with larger and longer volatility shocks than the ANN models. This result aligns with the results from Lim and Sek (2013),where an asymmetric GARCH model is found to be most preferred during a high volatility period.

The actual and forecasted values plotted in figures 6 and 7 for both models deliver further insight into these results. The ANN forecasted values track the actual volatility values more closely in the two low volatility sectors than the GARCH model. Both provide a similar range of forecasted values for the technology (medium volatility) and the manufacturing sector (high volatility), but the figures show the ANN model failing to accurately forecast volatility in the highest volatility sector, other, while the GARCH forecast was quite accurate. The learning curves in figure 5 provide evidence that this is not due to the ANN model overfitting the data. A possible reason for the ANN outperformance in the low volatility sectors and underperformance in the medium and high volatility sectors is due to the noise sensitivity of the backpropagation ANNs used in this study (Sharma, et al., 2017). In a lower volatility sector there is less noise in the data due to less price fluctuations, which the ANN can model and forecast quite accurately. In sectors with higher levels of volatility and therefore price fluctuations, the ANN loses accuracy when fitting the data and forecasting.

These results are subject to some limitations of this study that should be considered in future research. The most significant limitation is the small number of model-type specifications tested due to the document size constraints of this study. A future comprehensive study would include a greater number of GARCH-type models and ANN types and architectures. This would expand the evidence provided in this study to include an even greater guideline of model types to use and to which volatility profiles they should be applied for best forecasting results. Other factors that would expand upon this research include using larger timeframes (perhaps decades of return data) or using a different volatility proxy such as squared intra-day returns instead of squared daily returns.

Overall, this study demonstrates the applicability for ANN models to produce more accurate forecasts than the standard GARCH models when applied to specific volatility profiles. By examining the usefulness of the ANN model in comparison to the GARCH model, we have shown that the ANN parameters have the needed flexibility to improve performance. The GARCH models suffer from less flexibility as they are required to adhere to often strict constraints. Overall, the results are promising for ANN applications in volatility prediction, and with further refinement in the architecture performance can be improved.

# 7  Conclusion

The objective of this study is to evaluate the volatility prediction performance of an ANN model compared to the common GARCH-type models when applied to specific volatility profiles. This will identify specific volatility profiles where each model outperforms the other. Through this approach, the study will also demonstrate the usefulness of ANNs for the volatility prediction of assets within these specific volatility profiles. Three GARCH specifications and three ANN architectures are examined using the stock return data from 2005 to 2020 of all publicly traded companies in the United States. Each company is sorted into one of five sectors: consumer durables, health, technology, manufacturing, and other. A single best performing GARCH and ANN model is then chosen as the optimal model to move forward with the forecasting of volatility for each individual sector.

Based on the model AIC value, adherence to constraints, and other metrics, the EGARCH($p,q$) specification is chosen as the most adequate fit model over the GARCH($1,1$) and GARCH($p,q$) specifications for each sector. For consistency, the same approach is used for the ANN architecture selection, where the ANN(5,12,1) architecture is chosen over the ANN(5,1,1) and ANN(5,50,1) architectures based mainly on in-sample fit. By examining three distinct specifications, this approach helps ensure that the best GARCH specification and ANN architecture is chosen for volatility prediction performance comparisons.

The out-of-sample results clearly identify volatility profiles where each model outperforms the other. Using the RMSE metric as the main performance measuring criteria, the ANN model provides more accurate volatility forecasts for the low volatility sectors (consumer durables and health). This indicates that ANN models are a better choice than GARCH-type models when forecasting volatility in stocks with a low volatility profile. Meanwhile, the GARCH-type model outperformed the ANN model in the medium (technology) and high volatility (manufacturing and other) sectors. This indicates that GARCH-type models are more suitable and provide more accurate volatility forecasts for stocks with a medium or high volatility profile.

The results of this study have important implications in both industry and academics. By identifying the volatility profiles where each model will perform best, this study further solidifies the ANNs place in the volatility forecasters toolbox. The benefits of this study to industry is two-fold. First it gives economists, traders, and other market participants an additional volatility forecasting tool to the common GARCH model. Second, it provides evidence of the asset volatility profiles where each model type should be used. It is standard industry practice to never solely rely on a single type of model (Zhang, 2012). This study directly expands the types of models available to market participants. The study also furthers the academic literature in this area by expanding upon the existing work of volatility prediction models and adding further evaluation of the applicability of the ANN model in financial forecasting. Prior studies conducted focused on performance of ANN and GARCH models when applied to specific indexes and stocks, where this study has taken a broader view by grouping all stocks into volatility profiles and determining where each model performs best.

By using less complex ANN architectures and GARCH-type models, this study provides a foundation for further research in this context. Future research is required to examine more complex ANN architectures such as Long-Short Term Memory (LSTM) or Recurrent Neural Network (RNN), as well as GARCH-type models such as the TGARCH or GJR-GARCH. Broadening the set of models examined contributes to the literature by further identifying the specific scenarios where an ANN model or GARCH model should be used and gives market participants more choice of available models in both industry and academics.

# 8 References

*The Python code developed for this study and the datasets can be found at:*
*https://github.com/curtnybo/MSc-Dissertation-Code*

Abadi, M. et al., 2019. *TensorFlow White Papers - Large-Scale Machine Learning on Heterogeneous Distributed Systems.* [Online]
Available at: https://www.tensorflow.org/about/bib

Andersen, T. G. & Bollerslev, T., 1998. Answering the Skeptics: Yes, Standard Volatility Models do Provide Accurate Forecasts. *International Economic Review,* 39(4), pp. 885-905.

Arnerić, J., Poklepović, T. & Aljinović, Z., 2014. GARCH based artificial neural networks in forecasting conditional variance of stock returns. *CRORR Journal Regular Issue,* 5(2), pp. 329-343.

Bentes, S. R., 2015. Forecasting volatility in gold returns under the GARCH, IGARCH and FIGARCH frameworks: New evidence. *Physica A: Statistical Mechanics and its Applications,* 438(15), pp. 355-364.

Black, F., 1976. Studies of Stock Price Volatility Changes. *Proceedings of the 1976 Meetings of the Business and Economics Statistics Section American Statistical Association,* pp. 177-181.

Bollerslev, T., 1986. Generalized autoregressive conditional heteroskedasticity. *Journal of Econometrics,* 31(3), pp. 307-327.

Brooks, C., 2014. *Introductory Econometrics for Finance.* 3rd ed. Cambridge: Cambridge University Press.

Chong, C. W., Ahmad, M. I. & Abdullah, M. Y., 1999. Performance of GARCH models in forecasting stock market volatility. *Journal of Forecasting,* 18(5), pp. 333-343.

Ding, G. & Qin, L., 2020. Study on the prediction of stock price based on the associated network model of LSTM. *International Journal of Machine Learning and Cybernetics,* Volume 11, pp. 1307-1317.

Donaldson, G. R. & Kamstra, M., 1997. An artificial neural network-GARCH model for international stock return volatility. *Journal of Empirical Finance,* 4(1), pp. 17-46.

Engle, R., 2001. GARCH 101: The Use of ARCH/GARCH Models in Applied Econometrics. *Journal of Economic Perspectives,* 15(4), pp. 157-168.

Engle, R. F., 1982. Autoregressive Conditional Heteroscedasticity with Estimates of the Variance of United Kingdom Inflation. *Econometrica,* 50(4), pp. 987-1007.

Ezzat, H., 2012. The Application of GARCH and EGARCH in Modeling the Volatility of Daily Stock Returns During Massive Shocks: The Empirical Case of Egypt. *International Research Journal of Finance and Economics,* Volume 96, pp. 143-154.

French, K. R., 2020. *Kenneth R. French - Data Library.* [Online]
Available at: https://mba.tuck.dartmouth.edu/pages/faculty/ken.french/data_library.html
[Accessed June 2020].

Gabriel, A. S., 2012. Evaluating the Forecasting Performance of GARCH Models. Evidence from Romania. *Procedia - Social and Behavioral Sciences,* 62(24), pp. 1006-1010.

Géron, A., 2019. *Hands-On Machine Learning with Scikit-Learn, Keras, and TensorFlow.* 2nd ed. Sebastopol, CA: O'Reilly Media, Inc.

Goodfellow, I., Bengio, Y. & Courville, A., 2016. *Deep Learning.* s.l.:The MIT Press.

Hamid, S. A. & Iqbal, Z., 2004. Using neural networks for forecasting volatility of S&P 500 Index futures prices. *Journal of Business Research,* 57(10), pp. 1116-1125.

Heaton, J., 2008. *Introduction to Neural Networks for Java.* 2nd ed. s.l.:Heaton Research, Inc.

Hossain, A., Zaman, F., Nasser, M. & Mufakhkharul Islam, M., 2009. Comparison of GARCH, Neural Network and Support Vector Machine in Financial Time Series Prediction. *Pattern Recognition and Machine Intelligence. PReMI 2009. Lecture Notes in Computer Science,* Volume 5909, pp. 597-602.

Kim, Y. & Enke, D., 2016. Using Neural Networks to Forecast Volatility for an Asset Allocation Strategy Based on the Target Volatility. *Procedia Computer Science,* Volume 95, pp. 281-286.

Kristjanpoller, W., Fadic, A. & Minutolo, M. C., 2014. Volatility forecast using hybrid Neural Network models. *Expert Systems with Applications,* 41(5), pp. 2437-2442.

Kristjanpoller, W. & Hernández, E. P., 2017. Volatility of main metals forecasted by a hybrid ANN-GARCH model with regressors. *Expert Systems with Applications,* 84(30), pp. 290-300.

Krol, R., 2014. Economic Policy Uncertainty and Exchange Rate Volatility. *International Finance,* 17(2), pp. 241-256.

Laily, V. O. N., Warsito, B. & Asih, M. D., 2018. Comparison of ARCH / GARCH model and Elman Recurrent Neural Network on data return of closing price stock. *Journal of Physics: Conference Series,* Volume 1025.

Lim, C. M. & Sek, S. K., 2013. Comparing the Performances of GARCH-type Models in Capturing the Stock Market Volatility in Malaysia. *Procedia Economics and Finance,* Volume 5, pp. 478-487.

Lin, Z., 2018. Modelling and forecasting the stock market volatility of SSE Composite Index using GARCH models. *Future Generation Computer Systems,* 79(3), pp. 960-972.

Liu, H., Li, R. & Yuan, J., 2018. Deposit insurance pricing under GARCH. *Finance Research Letters,* Volume 26, pp. 242-249.

Liu, L. & Zhang, T., 2015. Economic policy uncertainty and stock market volatility. *Finance Research Letters,* Volume 15, pp. 99-105.

Lu, X., Que, D. & Cao, G., 2016. Volatility Forecast Based on the Hybrid Artificial Neural Network and GARCH-type Models. *Procedia Computer Science,* Volume 91, pp. 1044-1049.

MathWorks, 2020. *EGARCH conditional variance time series model - MATLAB.* [Online]
Available at: https://www.mathworks.com/help/econ/egarch.html
[Accessed July 2020].

Miron, D. & Tudor, C., 2010. Asymmetric Conditional Volatility Models: Empirical Estimation and Comparison of Forecasting Accuracy. *Romanian Journal of Economic Forecasting,* Volume 3.

Naik, N., Mohan, B. R. & Jha, R. A., 2020. GARCH-Model Identification based on Performance of Information Criteria. *Procedia Computer Science,* Volume 171, pp. 1935-1942.

Nelson, D. B., 1991. Conditional Heteroskedasticity in Asset Returns: A New Approach. *Econometrica,* 59(2), pp. 347-370.

Nielsen, M. A., 2015. *Neural Networks and Deep Learning.* s.l.:Determination Press.

Rumelhart, D. E., Hinton, G. E. & Williams, R. J., 1986. Learning Internal Representations by Error Propagation. In: *Parallel Distributed Processing: Explorations in the Microstructure of Cognition: Foundations.* s.l.:MIT Press, pp. 318-362.

Sharma, A., Bhuriya, D. & Singh, U., 2017. *Survey of stock market prediction using machine learning approach.* Coimbatore, India, IEEE.

Sheppard, K. et al., 2019. *ARCH Python - Release 4.13.* [Online]
Available at: https://github.com/bashtage/arch/tree/4.13

Wei, Y., Wang, Y. & Huang, D., 2010. Forecasting crude oil market volatility: Further evidence using GARCH-class models. *Energy Economics,* 32(6), pp. 1477-1484.

Yahoo Finance, 2020. *Yahoo Finance - Stock Market, Live Quotes, Business and Finance News - Historical Prices.* [Online]
Available at: https://ca.finance.yahoo.com/
[Accessed April 2020].

Zhang, P. G., 2012. Neural Networks for Time-Series Forecasting. In: *Handbook of Natural Computing.* Berlin, Heidelberg: Springer, pp. 461-477.